\documentclass[aps,prl,preprint,showpacs,preprintnumbers,amsmath,amssymb,groupedaddress]{revtex4-1}

\usepackage{soul}

\usepackage[dvipdfmx]{graphicx}
\usepackage{dcolumn}
\usepackage{bm}
\usepackage{color}
\definecolor{Red}  {rgb}{1,0,0}
\definecolor{Green}{rgb}{0,1,0}
\definecolor{Blue} {rgb}{0,0,1}
\usepackage{color}
\usepackage{umoline}
\usepackage{proofread}
\usepackage{lineno}

\newcommand {\bfv}[1] {{\boldsymbol {#1}}}

\newcommand\Rey{\mbox{\textit{Re}}}  
\newcommand\removed[1]{}

\usepackage{ulem}

\begin{document}

\title{Numerical reproduction of the spiral wave visualized experimentally in a wide-gap spherical Couette flow}


\author{}
\affiliation{}

\author{Kazuki Yoshikawa}
\affiliation{
  Department of Pure and Applied Physics, 
  Faculty of Engineering Science, Kansai University, Osaka, 564-8680, Japan
}
\author{Tomoaki Itano}
\email{itano@kansai-u.ac.jp}
\affiliation{
  Department of Pure and Applied Physics, 
  Faculty of Engineering Science, Kansai University, Osaka, 564-8680, Japan
}
\author{Masako Sugihara-Seki}
\altaffiliation{
  Graduate School of Engineering Science,
  Osaka University, Osaka 560-8531, Japan
}
\affiliation{
  Department of Pure and Applied Physics, 
  Faculty of Engineering Science, Kansai University, Osaka, 564-8680, Japan
}


\date{\today}

\begin{abstract}
SCF experiments were conducted according to the work of Egbers and Rath [Acta Mech. 111 pp. 125--140 (1995)].
Through visualization using aluminium flakes drifting on a horizontal plane illuminated by a laser sheet, the flow was identified as a spiral wave with azimuthal wavenumber $m=3$, using the experimentally obtained and numerically deduced comparison between phase velocities.
  By solving the equation of motion for the infinitesimal planar particles advecting in the flow field of the spiral wave, a visual distribution of reflected light was reproduced virtually, which is in good agreement with the picture obtained experimentally.
\end{abstract}

\pacs{
  47.27.De  
  47.20.Ky,  
  47.20.Qr,  
  47.32.Ef   
}


\maketitle

%
%
%
%

\section{Introduction}
The Newtonian fluid flow between double concentric spherical boundaries is a model in astronomical bodies \cite{Fow04,Feu11}.
Mechanical factors such as the Coriolis force, thermal instability due to gravity toward the centre of the system \cite{Cha61,Bus75,Zeb83}, or Lorentz force via the electromagnetic field \cite{Kid97,Sak99} have been incorporated into the governing equations.
In contrast, this study considers the flow transition triggered by angular differential velocities between the inner and outer spheres; the inner sphere rotates at a constant angular velocity and the outer sphere is at rest.
This flow, called the {\it spherical Couette flow} (SCF), could be a model of the liquid outer core of planetary and satellite bodies \cite{Son96} with different rotations of the inner core and outer shell if the Coriolis and magnetohydrodynamic effect would be negligible with their slow spins.

Early experimental studies using several different combinations of spherical boundaries with various radii \cite{Mun75,Bel84} suggested that SCF is slightly more complicated than cylindrical Taylor--Couette flow despite the apparent similarities.
Belyaef et al. \cite{Bel84,Bel91} estimated the first transitional Reynolds number for a relatively wide spherical radius ratio $\eta=r_{\rm in}/r_{\rm out}$ equal to $1/2$ using the power spectra obtained from laser Doppler velocimetry.
Using classical flow visualization techniques, Egbers and Rath \cite{Egb95} revealed a qualitative phase diagram of flow in SCF for $\eta=2/3$ and $3/4$.
The first transition from the laminar state for these radius ratios is triggered by a travelling sinusoidal disturbance at mid-latitudes propagating at a significantly low angular phase velocity in the azimuthal direction; it is displaced in an alternatively staggered way with respect to the equator.
The disturbance, known as a `spiral wave' \cite{Egb95}, can be visualized as a spiral pattern with $m$ equally spaced arms extending from the poles to the equatorial zone in each hemisphere; it is also observed in the flow on a rotating planar disk in a stationary casing \cite{Nak02}.
The spiral wave caused by crossflow instability for a relatively wide-gap SCF \cite{Nak02} is associated with an  inertial wave generated by an instability of the axisymmetric flows via Coriolis effect from a geophysical viewpoint.
Recently, the regimes of equatorial jet and Stewartson shear layer instabilities have been numerically and experimentally explored by further increasing the boundary rotation rate \cite{Fin12,Wic14,Bar18}.
The instabilities could also generate coherence around the poles, similar to the regular polygonal patterns formed in the polar jet stream on Jupiter and Saturn that have been observed in recent spacecraft missions \cite{God88}.

Improvements in computational capabilities have made it possible to numerically reproduce non-axisymmetric flows, such as spiral waves and turbulent transitions \cite{Dum94,Ara97,Hol06}.
The numerical linear stability analysis of the axisymmetric basic flow in the SCF can determine  the value of the critical Reynolds number $\Rey_{\rm cr}(m)$ for some representative values of $\eta$, over which a spiral wave with wave number $m$ supercritically bifurcates from the basic flow.
Additionally, competition between spiral waves with different wave numbers was observed on the route to turbulence \cite{Bel91,Abb18a,Abb18b,Got21}.
Using laser Doppler velocimetry, Wulf et al. \cite{Wul99} described flow transition over a critical $\Rey$ in a reconstructed state space.
According to them, the transition process may be described by a torus with modulation typical in the route to chaos via Hopf bifurcations and mode changes in the state space.

Flow visualization techniques enable us to identify transitions between the states of flow in SCF.
The measurement of flow in an SCF using laser Doppler velocimetry \cite{Bel84,Wul99}, which can provide a continuous time series of flow velocity at an observation point located in the domain, does not necessarily allow us to identify the state of flow describing the entire domain of the SCF.
  Particle image velocimetry, which was applied to experiments verifying the asymmetry of flow in either positive or negative Rossby numbers \cite{Hof19}, enables the identification of subtle differences among similar states of SCF.
  Although it may be difficult to distinguish the different states of flow emerging in the transition of SCF \cite{Egb95,Wul99,Nak02,Abb18a}, except for recognizing the difference in the wave number, a classical flow visualization technique using a mixed small amount of aluminium flakes is also still effective for identifying the first transition.
  While the value of the critical Reynolds number obtained by such a classical visualization technique is in good agreement with the numerical prediction \cite{Jun00}, it remains a question how a spiral wave bifurcating over the critical Reynolds number can be visualized even by a classical flow visualization technique like the mixing of a small amount of aluminium flakes to the working fluid.
Here, the objective of this study is to numerically reproduce the spiral wave of SCF by modelling flake motion advecting in the flow field.

The remainder of this paper is organized as follows.
The next section describes the SCF experimental apparatus used in this study with $\eta=1/2$ and a visualization configuration.
In Section 3, the dimensionless frequency is experimentally obtained and compared to that obtained from previous numerical calculations.
In Section 4, the equations of both translational and rotational motions of the infinitesimal planar particle drifting in flow are proposed.
Subsequently, the visualized image expected from the numerical calculations is presented.
The latter part of the section contains brief remarks on the difference between the experimentally and numerically visualized images.
Finally, Section 5 provides a summary of the study.

\section{Experimental Setup}
A gap between concentric double spherical boundaries having diameters of 85 mm and 170 mm was filled with a typical 22$\sim$36 wt\% aqueous glycerol solution mixed with a small amount of aluminium flakes (Daiwa Metal Powder Co., No. 1112, average particle diameter of 23 $\mu$m) and surfactant.
The outer sphere (container) was made of acrylic glass with optical access, except for the equatorial zone, which allows visualization of the flow in the gap.
The inner sphere, made of black anodized aluminium with its centre fixed at the origin, was suspended by the lower end of a stainless shaft located along the vertical $z$-axis. The angular velocity of the shaft was electrically controlled by a brushless DC motor (Oriental Motor, GFS2G5, BXM230-GFS, BXSD30-A).
The diameter of the shaft was 6 mm  with the intention being to not interfere with flow in the polar region.

A sphere rotating at a constant angular velocity $\Omega_{\rm in}=2\pi/T_{\rm in}$ produces a basic Stokesian laminar flow along spherical surfaces involving secondary circulation, i.e., Eckman downwellings at poles.
Interpreted in terms of mechanics, the meridional secondary flow streaming from the inner to outer sphere on the equator (or from the outer to inner sphere near the poles) is caused by the centrifugal force at the equator around the inner sphere under non-slip conditions owing to viscosity.

This axisymmetric basic flow becomes stronger as the Reynolds number $\Rey=r_{\rm in}^2\Omega_{\rm in}/\nu$ increases.
For $\eta=1/2$, the basic flow becomes unstable against infinitesimal sinusoidal disturbances with azimuthal wave number $m=4$, at $\Rey(m=4)=489$ \cite{Dum94,Jun00}.
However, another infinitesimal disturbance with wave number $m=3$, which is more dominant at significantly higher Reynolds numbers \cite{Got21}, may cause the $m=3$ spiral wave, which is often observed in experiments.
Belyaef et al. \cite{Bel84,Bel91} reported hysteresis loops related to the spiral waves with different wave number in multiple ranges of $\Rey$.
Such hysteresis among the spiral waves with $m=5,4,3$ has also been reported in another numerical study \cite{Abb18a} that examined flow at $\eta=2/3$.
Egbers and Rath \cite{Egb95} suggested that the hysteresis stems from the acceleration rate in prehistory of the flow development until the initial condition. However, we shall not delve into the depths of that in the present study.

Our experiments are designed as iterations of typical routines, described as follows.
First, the fluid temperature was measured when the inner sphere was at rest. Subsequently, the inner sphere was rotated at a constant angular velocity by initiating an abrupt change from the state of rest.
The inner sphere was rotated for more than 40 min, following which the fluid flow and its visualized appearance was recorded by a generic video camera with $1920 \times 1080$ pixels at a frame rate of 24 fps for approximately 5 min.
Keeping the inner sphere at rest for 5 min, we measured the fluid temperature again to ensure that the temperature difference from the initial value was within 0.5 $^\circ$C.
From the viscosity of the aqueous glycerol solution obtained using the fluid temperature \cite{Nia08}, the Reynolds number in the present experiments varied in the range of $500 \lesssim \Rey \lesssim 750$.
Here, note that if SCF is governed by the Navier--Stokes equations for incompressible flow with only the two control parameters, $\Rey$ and $\eta$, the state of SCF is determined by the initial condition from the viewpoint of the deterministic dynamical system.
If the outer sphere would rotate, it would be necessary to introduce the Rossby number as another dimensional parameter under the co-rotating frame.

\begin{figure}[h]
  \centering
  \includegraphics[angle=0,width=0.75\columnwidth]{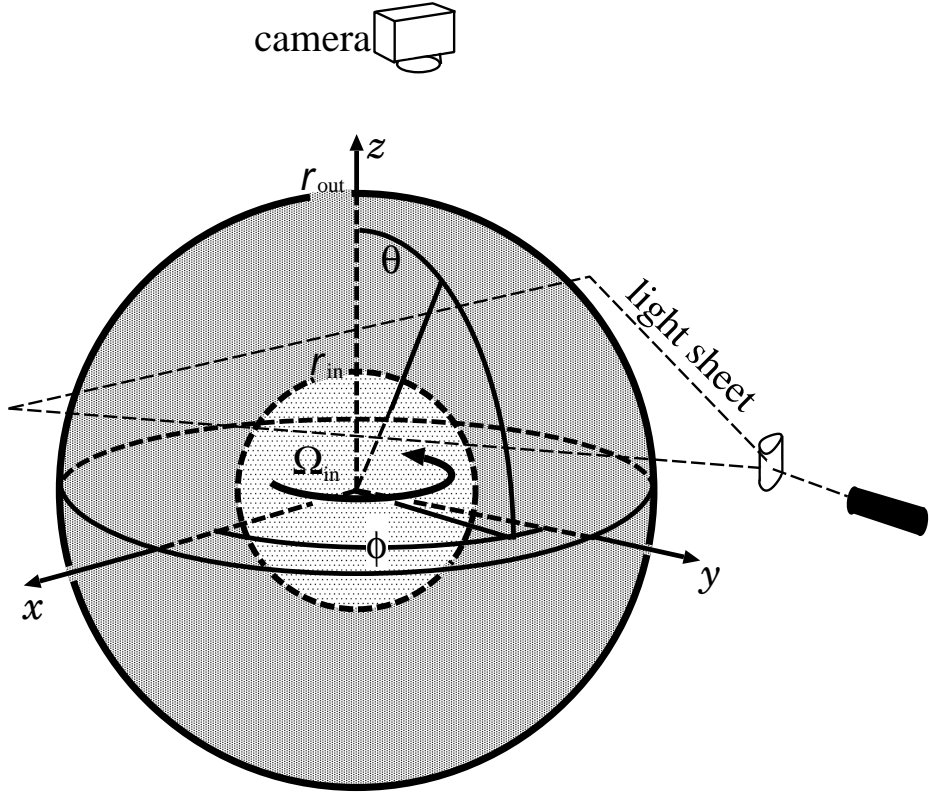}
  \caption{
    Schematic of the experimental setup.
    The incompressible fluid is confined between the inner and outer spheres with radii $r_{\rm in}$ and $r_{\rm out}$, respectively.
    The inner sphere rotates at a constant angular velocity, $\Omega_{\rm in}$, with respect to the $z$-axis.
    The nondimensional geometrical parameter is either the gap ratio $\beta=(r_{\rm out}-r_{\rm in})/r_{\rm in}$ or aspect ratio $\eta=r_{\rm in}/r_{\rm out}$
  }
  \label{fig.config}
\end{figure}

Fig.\ref{fig.config} illustrates the setup of a light source and camera in the present experiment.
To illuminate the flow, a laser beam of H1.6 mm $\times$ W0.9 mm in diameter was emitted parallel to the $y$-axis from the light source (Integrated Optics, 0520 L-11A, CW 520 nm / 200 mW).
The beam was spread uniformly on a horizontal plane at $z=r_{\rm in}$ through a Powell lens located at $(x,y,z)=(0,y_0,r_{\rm in})$.
If most of the aluminium flakes, which drift at a location on the plane illuminated in the flow, are oriented in a specific direction owing to the shear of the flow, the reflection from the location toward the camera may intensify or diminish compared to the average.
The contrast pattern in the image captured by the camera reflects the shear structure of the flow experienced by the flakes.
Moreover, if the flow structure is axisymmetric, the light intensity at a fixed observation point will not vary with time.
However, if the flow structure is not axisymmetric, it will vary with time.

\begin{figure}[t]
  \begin{center}
    \vskip -8mm
     \includegraphics[scale=0.75]{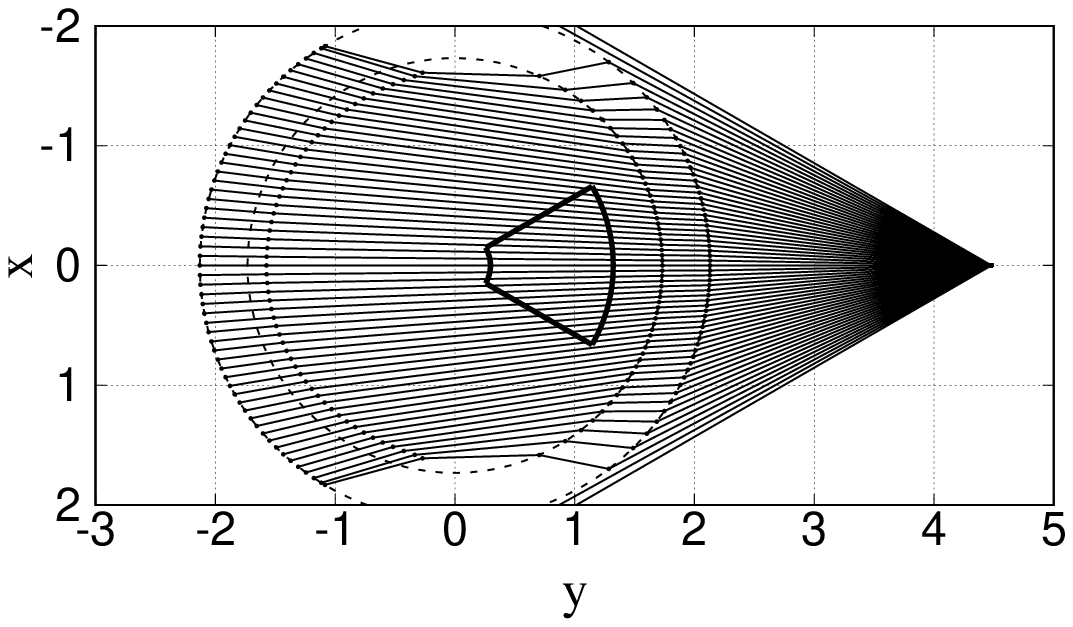}
    \vskip -25mm
     \includegraphics[scale=0.75]{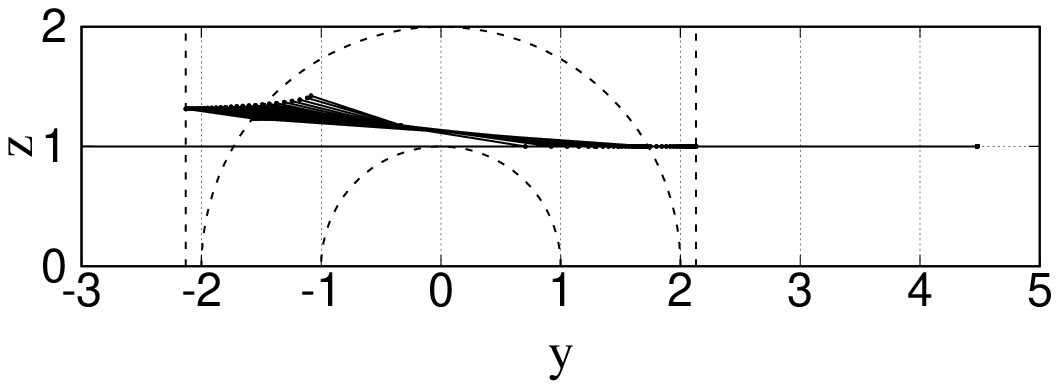}
    \vspace*{-20mm}
  \end{center}
  \caption{
    The light rays emitted from a Powell lens located at $y_0=190$ mm in case of $\eta=1/2$ are described horizontally every $1^{\circ}$.
    The upper panel shows the projection onto the $x-y$ plane, and the lower panel shows the projection onto the $y-z$ plane.
    The divisions on axes are scaled by the radius of the inner sphere, $r_{\rm in}=42.5$ mm.
    The dashed curves indicate the section of the container at the planes $z=r_0$ and $x=0$, respectively
  }
  \label{lens}
\end{figure}

The inner surface of the acrylic container is spherical and has a diameter of $170$ mm centred at the origin, and the outer surface of the container is cylindrical having a diameter of $181$ mm centred about the $z$-axis.
Without refractive index matching, the acrylic container refracts the laser sheet at the surface of the container against both ambient air and the working fluid.
However, fortunately, the container with spherical inner and circular outer surfaces partly acts as a collimator lens for the present light sheet by adjusting the distance to the lens, $y_0$.
By some adjustment, the rays across the fluid domain approach approximately parallel to the $y$-axis, and the spatial distribution of the light intensity in the domain may be expected to be relatively uniform.
Fig. \ref{lens} illustrates the light rays scattered by the lens and refracted at the boundaries, which were obtained from the calculation based on the refractive indices of the air, container, and fluid as $n=1.00, 1.49,$ and $1.33$, respectively.
The ray tracing used in the figure is a simple geometrical mapping, where refraction, including total internal reflection, was simulated by extracting the correct refraction point and angle.
The uniformity of light intensity in an image of flow visualization may be ensured if reflective flakes are uniformly distributed in a domain with isotropic orientation \cite{Got11}.

\section{Spiral wave}
Suppose that a spiral wave with $m$ arms extending equally from the poles to the equator continues to rotate around the $z$-axis at a constant angular velocity $\omega$ without changing its shape.
The flow structure of the spiral wave satisfies the following symmetry:
\begin{eqnarray*}
  u_r     (r,\pi-\theta,\phi+\pi/m) &=&  u_r     (r,\theta,\phi) \ \ , \\
  u_\theta(r,\pi-\theta,\phi+\pi/m) &=& -u_\theta(r,\theta,\phi) \ \ , \\
  u_\phi  (r,\pi-\theta,\phi+\pi/m) &=&  u_\phi  (r,\theta,\phi) \ \ ,
\end{eqnarray*}
as well as the periodicity with wave number $m$ in the azimuthal direction $u_i(r,\theta,\phi+2\pi/m)=u_i(r,\theta,\phi)$ ($i=r,\theta,\phi$). 
The symmetry requires that the flow patterns appearing in both hemispheres shift toward each other by half the azimuthal wavelength with the reflection at the equatorial plane.
High-shear regions localized around the arms of the spiral wave rotate around the $z$-axis such that the light intensity in a narrow area in the mid-latitude zone, which is captured by the camera fixed at $z \gg r_{\rm out}$, may oscillate periodically at intervals when an arm passes through the area.
Given that the spiral wave satisfying the $m$-fold symmetries is considered as an attractor bifurcated from the axisymmetric basic flow via nonlinearity in the SCF dynamical system, it is hereafter referred to as a spiral {\it state}.

\begin{figure}[h]
  \centerline{
    \includegraphics[angle=0,width=1.05\columnwidth]{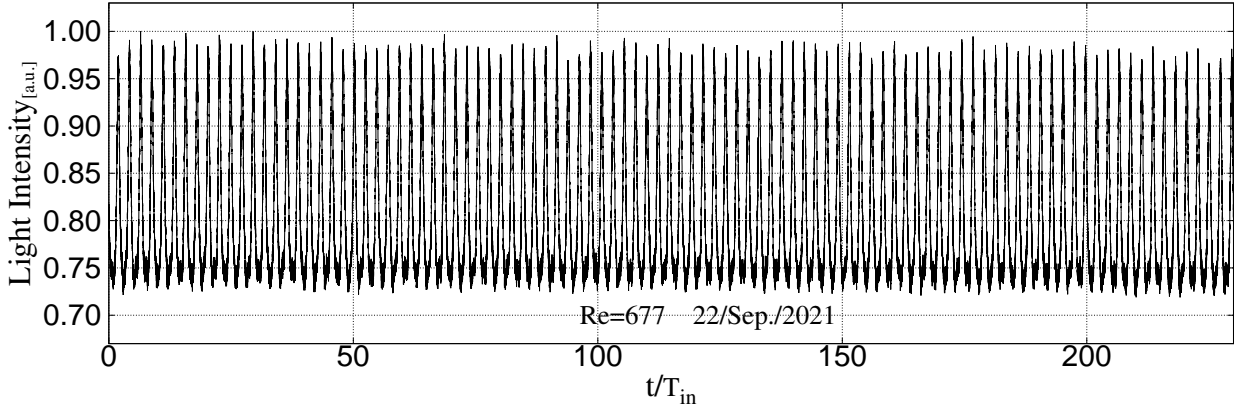}
  }
  \caption{
    Time series of light intensity in an area in the mid-latitude zone.
    $\eta=1/2$, $T_{\rm in}=12$ sec, $\Rey=677$
  }
  \label{light intensity}
\end{figure}

In the actual experiment, we recorded the time series of the light intensity in a narrow area, which is surrounded by thick solid lines in Fig.\ref{lens}, on the illuminated horizontal plane.
The area shaped as a sector centred on the $y$-axis spans the range $|\phi-\pi/2| < \pi/6$ in the azimuthal direction.
Fig.\ref{light intensity} shows an example of the oscillation of the recorded light intensity, which was used to calculate the period $T$, via Fourier transform.
This period is related to the wave number $m$ and angular velocity of the spiral state $\omega$ as $m\omega=2\pi/T$.

Note that the values of $m$ and $\omega$ cannot be independently estimated in principle until the spiral state is acquired in the entire domain; however, the product can be obtained directly from the time series.
Here, we split the sector depicted in Fig.\ref{lens} into two segments ($x>0$ and $x<0$), and measured a certain phase lag $\Delta t$ from two time series of light intensity obtained at the segments.
The angular velocity $\omega$ can be calculated from $\omega=\Delta \phi/\Delta t$, where $\Delta \phi=(\pi/6)/2$.
Therefore, from the measurements of $T$ and $\omega$, we could obtain the wave number $m = 3.11\pm 0.15$ for the spiral state in the present study.

\begin{figure}[h]
  \centerline{
    \includegraphics[angle=0,width=1.15\columnwidth]{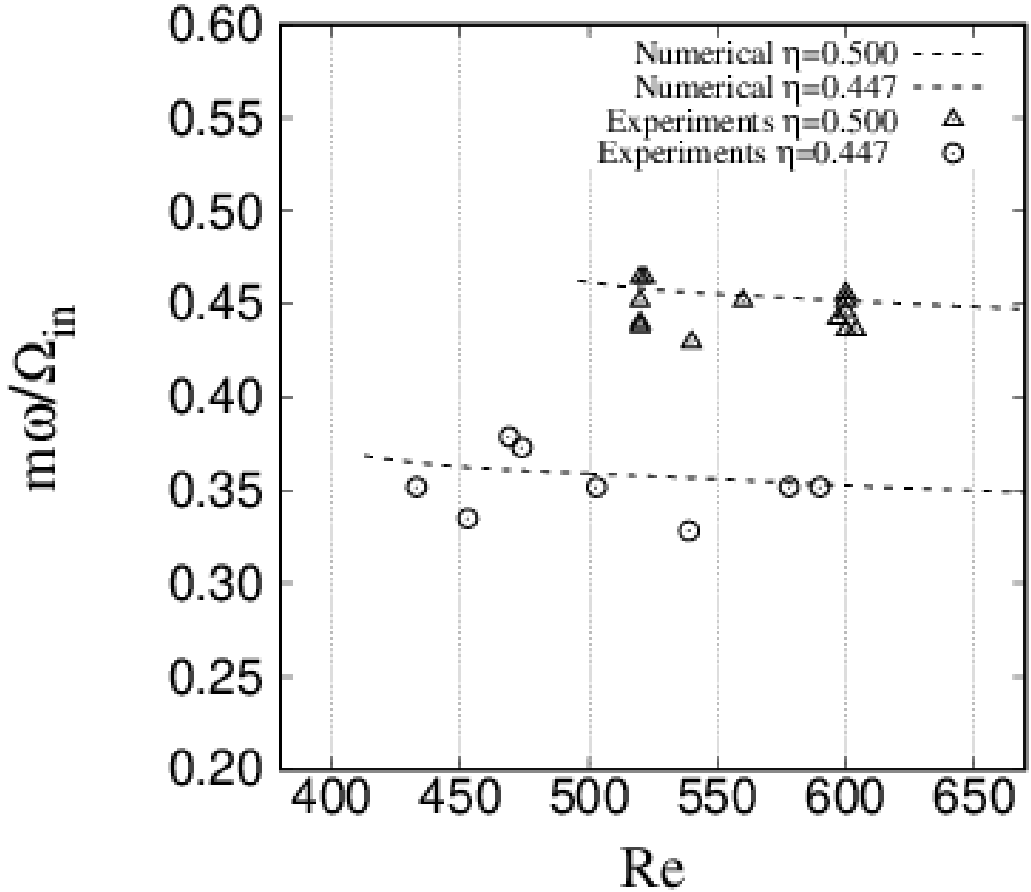}
  }
  \caption{
      The dimensionless frequencies against $\Rey$ for $\eta=1/2$ ($\triangle$) and $\eta=0.447$ ($\circ$).
    The dimensionless frequency is equivalent to $T_{\rm in}/T$, where $T_{\rm in}$ is the rotation period of the inner sphere. The period $T$ is calculated from the time series of reflection light intensity in an area in the mid-latitude zone.
    Dashed curves are based on the phase angular velocities $\omega$ of spiral states numerically solved with the $3$-fold symmetry for $\eta=1/2$ (upper) and $0.447$ (lower) \cite{Got21}
  }
  \label{frequency}
\end{figure}

The experiments were conducted using two inner spheres with different radii, $r_{\rm in}=42.5$ mm and $r_{\rm in}=38.0$ mm, which  correspond to $\eta=1/2$ and $\eta=0.447$, respectively, for the present container with radius $r_{\rm out}=85.0$ mm.
Here, we define the product of $m$ and $\omega$, nondimensionalized by $\Omega_{\rm in}$, as the {\it dimensionless frequency} in accordance with Ref. \cite{Bel91}.
Fig.\ref{frequency} shows the dimensionless frequency against $\Rey$ obtained from the present experiment with $\eta=1/2$ and $\eta=0.447$ against $\Rey$.
The range of the Reynolds numbers was $500 \lesssim Re \lesssim 750$ for each $\eta$.

Direct numerical simulations of SCF satisfying the Navier--Stokes equations were performed recently \cite{Ina19}.
The flow field was expanded into a series of spherical harmonics and modified Chebyshev polynomials, as used in previous numerical studies \cite{Sch13,Fri05,Ita09}, and the Helmholtz equation equivalent to the Navier--Stokes equation was solved numerically with the aid of LAPACK libraries \cite{And99}.
Moreover, using some approximately equilibrium states obtained from the simulations as seeds, the spiral states with $m=4, 3$, and $2$ were exactly solved using the Newton--Raphson algorithm; hence, the angular phase velocity was also specified numerically \cite{Got21}.
The dimensionless frequencies calculated from the spiral states with $m=3$ solved for $\eta=1/2$ and $0.447$ are plotted as dashed curves in Fig.\ref{frequency} for reference.
Although not shown in the figure, those of the spiral states with $m=4$ and $2$ were in the range of $0.6\pm 0.03$ and $0.28\pm 0.02$, respectively, for $\Rey\le 700$, which suggests that the experimentally obtained flow is the spiral state with $m=3$.
For reference, Ref. \cite{Bel91} reported 0.614 as a typical value of the dimensionless frequency of a sinusoidal perturbation with the azimuthal wave number $m=4$.
A comparison of $m=3$ and $m=4,2$ in dimensionless frequency suggests that the present state that was realized experimentally corresponds to the spiral state with $m=3$, which remains within a relative error of 8\% of its value for various Reynolds numbers.

A spiral state with $\eta=0.447$ emerges over $\Rey_{\rm cr} \approx 410$ smaller than that with $\eta=1/2$.
The fact that $\Rey_{\rm cr}$ increases with increasing $\eta$ is in qualitative agreement with the phase diagram reported in Ref. \cite{Egb95}.
The dimensionless frequency for both cases is kept almost constant; however, it decreases slightly with an increase in $\Rey$, which is in good agreement qualitatively with Ref. \cite{Bel91}.
In addition, the figure shows that the dimensionless frequency for $\eta=0.447$ is smaller than that for $\eta=1/2$.
An idealized spiral state may be regarded as an instability of the Stewartson shear layer compensating for the angular velocity difference between the inner and outer boundaries \cite{Hof19}.


\begin{figure}[h]
  \centerline{
    \includegraphics[angle=0,width=0.35\columnwidth]{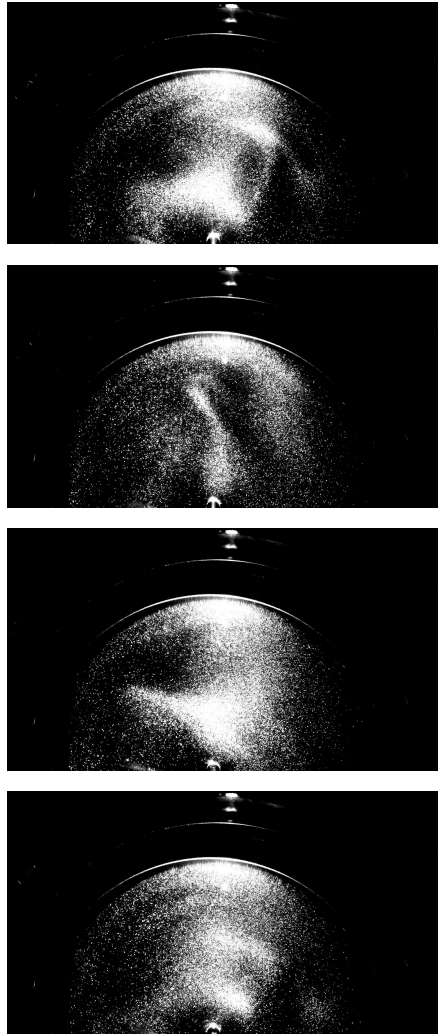}
  }
  \caption{
    Sequential snapshots of the flow at $\eta=1/2$ visualized by aluminium flakes drifting on the approximately horizontal plane $z=r_0$ illuminated as shown in Fig. \ref{lens}.
    The Reynolds number $\Rey=604$ was estimated using the concentration of the glycerol solution and fluid temperature.
    The practically obtained interval of snapshots was 10 s, which is close to the numerically deduced value, $T/9=9.6$ s, from dimensionless frequency at $\Rey=604$.
    The top of the inner sphere and the shaft along the $z$-axis are observed at the centre below these snapshots, and the sphere rotates counter-clockwise around the axis.
  }
  \label{fig:real}
\end{figure}

Fig.\ref{fig:real} shows four sequential snapshots of the flow visualized by the aluminium flakes that reflect light rays on the illuminated approximately horizontal halfplane, $z\approx r_0$ and $y \gtrsim 0$, as shown in Fig.\ref{lens}.
The zenith of the inner sphere and shaft are seen at the centre below these snapshots, and the sphere rotates counter-clockwise around the axis.
In these snapshots, the pixel size was about 0.096 mm and the frame rate was 24 s$^{-1}$.
The surface of the inner sphere at the equatorial region is estimated to move within $0.85$ mm at most by a frame, which ensures the temporal resolution.
Thus, blurred stripes as if scrubbed towards the azimuthal direction in the figures are probably peculiar to the spiral states of SCF.
The wave number of the state of the flow cannot be estimated based on a quick glance at these snapshots. Thus, the above discussion on dimensionless frequency may ensure that the state of the flow is in a spiral state with wave number $m=3$.
The intervals of the sequential snapshots were adjusted only to $T/9$, such that four snapshots can describe only a single period.
Here, assuming $m=3$, the period $T=2\pi/\omega$ was estimated from the numerically obtained value of the dimensionless frequency at $\Rey$ calculated from the fluid temperature and glycerol solution concentration used in the experiment.
The numerically estimated dimensionless frequency at $\Rey=604$ was $0.452$, and thus, $T/9$ equals to 9.64 s for the rotation rate 4.61 rpm adopted in the present experiment.
In Fig.\ref{fig:real}, the practically obtained interval of snapshots selected to complete a single period of recorded time series was 10 s.
With some exceptions, the fourth snapshot is similar to the first, which satisfies the periodic condition in the azimuthal direction of the spiral state with wave number $m=3$.
The difference between them might be associated with symmetry breaking, which would occur with successive phase transitions above the critical Reynolds numbers.
%
%

A bright pattern shaped as the constricted neck of a crane extending from the shaft was observed along the $y$-axis in the second snapshot of Fig.\ref{fig:real}.
The exact spiral state is frozen in the rotation with a constant angular velocity around the $z$-axis, i.e., rotating without any change in its flow structure over time; thus, one would expect that the neck captured at the sequential snapshot should rotate forward to the azimuthal direction by $2 \pi/9$ without any change in pattern.
However, the neck appeared more constricted at the second snapshot than at the third snapshot.
Similarly, in the first quadrant, any copy of the neck appears neither in the first nor in the fourth snapshot; instead, blurred shadow regions in the shape as a horn extends in the direction $\phi=\pi/4$ in the first quadrant in both snapshots.
This was incorporated with other shadow fragments into a shadow region adjacent to the crane's neck in the second snapshot.
Even if the flow structure and orientations of the drifting flakes are frozen in rotation by just $2\pi/9$, the direction of a light ray reflected by flakes drifting on the plane is not kept constant over time.
This does not occur when the optical axes of both the light source and camera accord by means of a half-mirror, which was employed in Ref. \cite{Egb95}.

\section{Numerical model of flake motion}
In the experiments conducted in this study, aluminium flakes were mixed with an aqueous glycerol solution, and they diffused and advected in the flow by changing their orientations according to the shear velocity gradient.
According to Goto et al. \cite{Got11}, if the size of the flake is much smaller than the characteristic length of fluid motion and if the Stokes number is much less than unity, the contrast in light intensity of visualization images of the flow mixed with reflective flakes stems from the non-isotropic orientations of the flakes rather than their spatially non-uniform accumulation.
The motion of the flakes in the flow can be assumed to be the rotation of infinitesimal planar particles without inertia advecting passively in the flow.
We solved the simultaneous equations of both the translational and rotational motions of the particle drifting in a spiral state flow corresponding to $m=3$.
The temporal evolution of a particle located at $\bfv{x}(\tau)$ with orientation $\bfv{n}(\tau)$ at time $\tau$ is as follows.

Suppose that the particles are uniformly distributed on the illuminated plane $z=r_0$ at time $t$ between the spherical boundaries, where the $i$-th particle is located at $\bfv{x}_i(t)$.
With the aid of utilities in the numerical libraries of spherical harmonics \cite{Sch13,Fri05}, the position of the $i$-th particle at $t-T_0$ can be solved by integrating the following translational motion equation backward in time:
 $\displaystyle \frac{d\bfv{x}_i}{d\tau}=\bfv{u}(\bfv{x}_i,\tau)$.
Thus, the map from a slightly complicated curved surface consisting of a set of $\bfv{x}_i(t-T_0)$ to $z=r_0$ is the translation of particles by a time advance from $t-T_0$ to $t$.

Next, each particle at the instance of $t-T_0$ was assumed to face the orientations of the 12 vertices of the icosahedron with respect to its centre with an equal probability.
The governing equation for the orientation of each particle \cite{Got11} is $\displaystyle \frac{d\bfv{n}_i}{d\tau}=\bfv{n}_i\times\bfv{n}_i\times\bfv{\nabla}(\bfv{n}_i\cdot\bfv{u})$, where the unit normal vector of the face at instance $\tau$ is denoted by $\bfv{n}_i(\tau)$, and the velocity field at $\bfv{x}=\bfv{x}_i(\tau)$ is applied to the value of $\bfv{u}$ in the equation.
Note that the divergence of the right-hand side of the equation by $\bfv{n}$ does not necessarily vanish, and hence, the probability of the orientation on the unit sphere may either accumulate or diffuse with time.
To obtain the velocity field at $\tau$, the velocity field solved using the Newton–-Raphson method was rotated numerically by $\omega \tau$ around the $z$-axis.
We determine the orientation of the $i$-th particle at the original instance $t$ by integrating the governing equation of the orientation forward in time until each particle arrives at the illuminated plane $z=r_0$.
If the particles are strongly sheared during the integrated period, they can be aligned in a particular direction at $t$, even if the plane particles are oriented isotropically at $t-T_0$.
Since $\bfv{n}_i(\tau)$ moves on the unit sphere with time, the governing equation of the particle rotation was converted to the equivalent first-order Euler method of a quaternion algorithm, which was numerically integrated with a fine time step.
Here, we adopted $2.8 T_{\rm in}$ as $T_0$ under the expectation that the order of $T_{\rm in}$ is sufficiently long for particles oriented isotropically at the initial stage to experience the shear region of the spiral state.
For reference, $2.8T_{\rm in}$ is 36 s for the rotation rate 4.61 rpm for Fig.\ref{fig:real} .

The direction of the incident rays $\bfv{I}$ to a particle is assumed to be $\bfv{I}=-\bfv{e}_y$, independent of the position of the particle.
If the orientation of a particle $\bfv{n}$ is in the bisectional direction between $-\bfv{I}$ and the optical axis of the camera, reflected light is observable.
In general, the direction of reflected ray $\bfv{R}$ from the particle is $\bfv{R}=\bfv{I}-2(\bfv{I}\cdot\bfv{n})\bfv{n}$.
The alignment of the reflected ray against the optical axis of the camera must be correlated with the intensity distribution in the recorded images.
Thus, the degree of reflected light intensity captured by the camera can be evaluated by the magnitude of $\langle\bfv{C}\cdot\bfv{R}\rangle$, where $\bfv{C}$ is the unit vector along the optical axis of the camera lens with respect to the reflecting particle, and we assumed $\bfv{C}=\bfv{e}_z$ for ease.
Here, the square bracket represents the ensemble average, such that the contrast in light intensity at $\bfv{x}$ is the average of $\bfv{C}\cdot\bfv{R}$ for all computed flakes positioned at $\bfv{x}$.
If the orientations of flakes at $\bfv{x}$ are distributed isotropically, the intensity vanishes; the orientation of the flakes at $\bfv{x}$ is aligned in the direction, and the intensity is close to $\pm 1$.

\begin{figure}[h]
  \centerline{
    \includegraphics[angle=0,width=0.35\columnwidth]{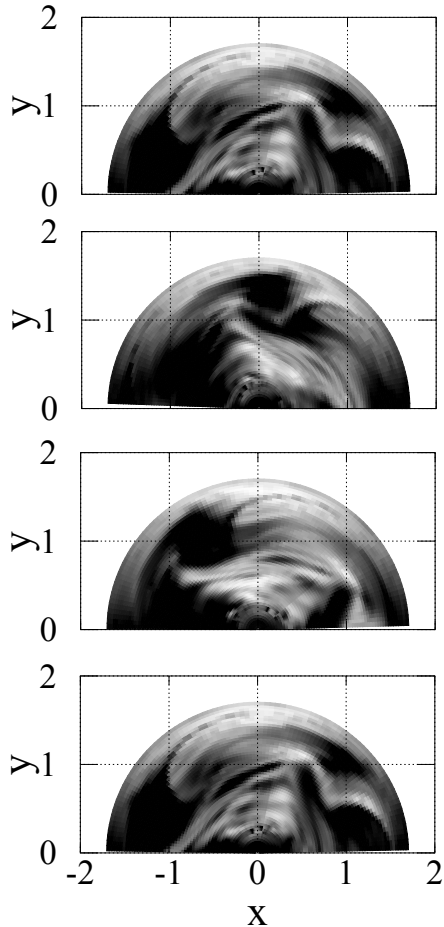}
  }
  \caption{
    Time series of contrast in light intensity reflected from infinitesimal planar particle on the plane $z=r_0$ advecting in the flow field of a spiral state of wave number $m=3$ at $(\eta,\Rey)=(1/2,560)$ was calculated.
    The divisions on axes are scaled according to the radius of the inner sphere, $r_{\rm in}$.
    The images correspond to Fig.\ref{fig:real}, $t=0$, $T/9$, $2T/9$, $3T/9$, from the top to the bottom, respectively
  }
  \label{fig:virtual}
\end{figure}

Fig.\ref{fig:virtual} shows a time series of the contrast pattern of light intensity calculated visually, which is generated by the reflection from infinitesimal planar particles advecting in a spiral state corresponding to $m=3$ at $\Rey=560$.
The interval between the figures was $T/9$ deduced from the dimensionless frequency at the corresponding Reynolds number.
The range of the contrast from $0$ to $1$ corresponds to the greyscale in Fig.\ref{fig:virtual}.
The value of the contrast is negative in most of the region $y<0$ for $\bfv{i}=-\bfv{e}_y$ (not shown in the figure), even though the incident ray is assumed to be distributed uniformly in the region $y<0$, as in the region $y>0$.
This was not confirmed in our experiment because light emitted from $y>0$ is practically absorbed by flakes floating in the region $y>0$.

A shadow region as a circular spot centred at $(x,y)=(-0.5,1)$ is identified in the third part of Fig.\ref{fig:virtual} ($t=2T/9$).
This shadow region is probably identical to that located around $(x,y)=(0.2,1.3)$ in the second part of Fig.\ref{fig:virtual} and around $(x,y)=(-1.2,0.5)$ in the fourth Fig.\ref{fig:virtual}.
A bright pattern at the right-hand side, adjacent to the spot, is reminiscent of the pattern shaped as a crane's constricted neck extending from the shaft, which was observed along the $y$-axis in the second snapshot of Fig.\ref{fig:real}.
Both the neck and shadow regions appear to change their shapes over time.
Although the $m=3$ spiral state is principally a simple rotating wave solution without its shape change, the contrast of the light intensity changes over time, which is confirmed both experimentally and numerically.
As shown above, there are similarities between Fig.\ref{fig:real} experimentally captured at $\Rey=604$ and Fig.\ref{fig:virtual} numerically calculated at $\Re=560$.
From this similarity, it seems possible to relate the experimental visualization images to the numerically obtained velocity field of the spiral state. 


\section{Discussion}
\begin{figure}[t]
  \centerline{
    \includegraphics[angle=0,width=0.35\columnwidth]{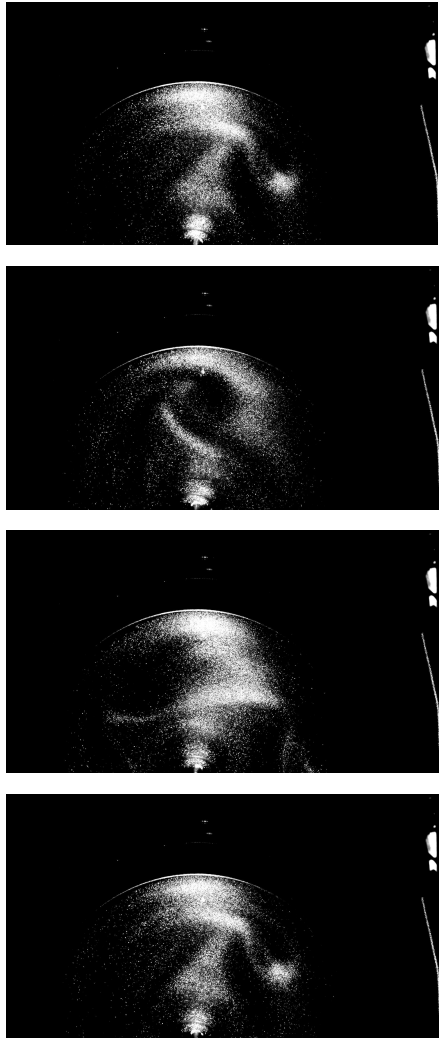}
  }
  \caption{
    Similar to Fig.\ref{fig:real} for $\Rey=751$ ($\eta=1/2$).
    The practical interval of snapshots was 4.75 s, which is close to the numerically deduced interval of $T/9=4.5$ s from dimensionless frequency at $\Rey=751$
  }
  \label{fig:real2}
\end{figure}

Fig.\ref{fig:real2} shows sequential snapshots of the flow realized at $\Rey=751$.
We did not observe any qualitative distinction in the visualization of spiral state in the Reynolds number range from 560 to 750.
We adopted $4.75$ s as the interval of snapshots in the figure, whereas numerical studies predicted the interval to be $T/9=4.5$ s.
In the first snapshot, blurred shadow regions in the shape of a horn extended in the $\phi=\pi/4$ direction in the first quadrant, which was incorporated with shadow fragments into a circular shadow region on the $y$-axis in the second snapshot, and then developed into a larger shadow region in the second quadrant in the third snapshot.
Note that this circular shadow should not be regarded as a swirling fluid motion until the numerically obtained flow field is compared to the snapshot.
The fourth snapshot is quite similar to the first snapshot, which suggests that the realized state satisfies the periodic condition of the spiral state with wave number $m=3$.
A flow state practically consists of combinations of a small set of incommensurable frequencies, and the spiral state obtained at $\Rey=751$ was complete as compared to that obtained at $\Rey=604$, which contained a certain modulation with lower frequencies \cite{Bel91}.

%
%

From a deterministic point of view, the initial conditions would influence the visualization results.
In the case of periodic flows, there is no trivial consensus as to what initial distribution of orientations of the flakes should be chosen.
However, our experimental visualization of the flow has shown that there is a unique spatial distribution of orientation preferred by the particles drifting in the spiral state of SCF.
In our numerical model, we assumed that the orientation of a flake is isotropically distributed at time $t-T_0$, where the flake locates at a location on the laser sheet at time $t$ of observation.
Although our calculations were performed by changing an arbitrary value $T_0$ from 1.4$T_{\rm in}$ to $5.6T_{\rm in}$, we obtained qualitatively similar patterns as in Fig.\ref{fig:virtual}.
Within the examined range of $T_0$, the difference of the initial condition hardly affects the similarity between the experimental visualization images and the numerically obtained image of the spiral state.

\begin{figure}[t]
  \centerline{
    \includegraphics[angle=0,width=0.50\columnwidth]{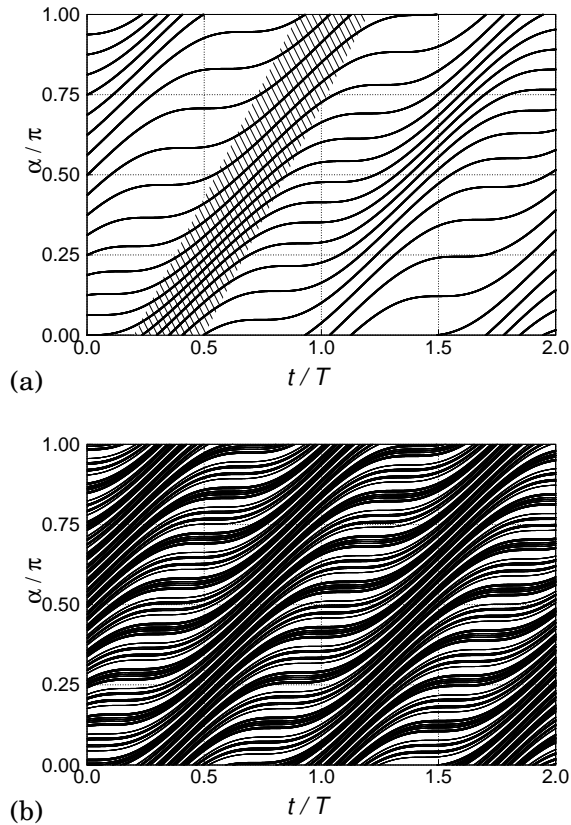}
  }
  \caption{
    The time evolutions of the orientation $\alpha$ for flakes in isotropically oriented at (a) $t=0$ and (b) $t=(n/10)T$ ($n=0,1,\cdots,9$). Flow function on a two-dimensional plane is specified from parameters $(\omega_0,\Omega_0)=(2\pi,\pi)$.
  }
  \label{fig:toy}
\end{figure}

Experiencing local shear in flow, particles change their orientation, meaning the distribution of particle orientation may change from isotropic to biased.
On the other hand, thermal factors such as the thermal fluctuation of working fluid, differences in the shape of individual particles with finite size, or collision of particles, which have not been taken into account in the present numerical model of particle orientation, are likely to make the distribution of their orientation isotropic.
By continuously initializing the particle orientations to be isotropic at multiple times, we could take the latter factor into the numerical model.

Consider a stream function on a two-dimensional plane, $\varPsi(\bfv{r})=-\Omega_0\bigl(\bfv{n}(t)\cdot\bfv{r}\bigr)^2/2$, where $\bfv{n}(t)=\bfv{e}_x\cos{(\omega_0 t)}+\bfv{e}_y\sin{(\omega_0 t)}$.
The parameters, $\Omega_0$ and $\omega_0$, correspond to the shear rate and the time variation of flow direction that a drifting flake experiences, respectively.
In the special case of $\omega_0=0$ the two-dimensional flow field $(u_x,u_y)=\bigl(\frac{\partial \varPsi}{\partial y},-\frac{\partial \varPsi}{\partial x}\bigr)$ corresponds to the simple Couette flow.
For $\omega_0(\omega_0-\Omega_0)>0$, the governing equation of the particle rotation, $\dot{\bfv{n}}_i=\bfv{n}_i\times\bfv{n}_i\times\bfv{\nabla}(\bfv{n}_i\cdot\bfv{u})$, has the general solution of particle orientation, $\bfv{n}_i=\bfv{e}_x\cos{\alpha}+\bfv{e}_y\sin{\alpha}$, where $\alpha=\omega_0 t+\arctan{\Bigl(-\frac{\omega_0}{\sigma}\tan{\sigma(t-t_{\rm i})}\Bigr)}$, $\sigma=\sqrt{\omega_0(\omega_0-\Omega_0)}$ and an arbitrary constant $t_{\rm i}$ is determined by the initial condition of particle orientation.
Fig.\ref{fig:toy}(a) shows an example of the time evolution of particle orientation, $\alpha$, which initially oriented to be isotropic at $t=0$.
A time evolution of orientation of a particle drifting in the flow is represented as a trajectory in the $t$-$\alpha$ plane.
The value of $\alpha-\omega_0 t$ of the exact solution are periodic with respect to period $T=\pi/\sigma$, because of no difference between the front and back surfaces of a particle.
A zone where trajectories are relatively dense is indicated as a shaded stripe zone in the figure.
If the values of $\alpha$ in the zone at a time $t$ is coincident with the angle between the incident ray of the laser sheet and the optical axis of the camera, light intensity reflected from particles is expected to be relatively intensified.

In general, the thermal factor continuously cause particle orientation to be isotropic.
Superimposing deterministic trajectories of particle orientation obtained from initializing particle orientation as isotropic at successive multiple times, we pseudo-reproduced the trajectories of particle orientation under the thermal factor in Fig.\ref{fig:toy}(b).
In the figure, the distribution of particle orientations is initialized to be isotropic at uniformly distributed successive times at $t=(n/10)T$ where $n$ is $0,1,\cdots,9$.
We can observe again several oblique zones where trajectories are relatively dense in the figure, one of which is coincident with the shaded stripe zone in Fig.\ref{fig:toy}(a).
This implies that exact trajectories of particle orientations calculated based on a deterministic view may virtually reproduce an experimental visual distribution of reflected light reflecting local shear in flow, even under an assumption that the distribution of particle orientation is initialized to isotropic at a time without taking continuous thermal factor into account.


\section{Summary}
This study conducted SCF experiments according to the work of Egbers and Rath [Acta Mech. 111 pp. 125--140 (1995)]\cite{Egb95}.
A comparison of the dimensionless frequency between the experimental and numerical results for $\eta=1/2$ and $\eta=0.447$ suggested that a spiral state with wave number $m=3$ was realized in our experiments.
The spiral states were visualized using aluminium flakes drifting on a horizontal plane illuminated by a laser sheet.
Solving the equations of motion on translation and orientation for the infinitesimal planar particles advecting in the flow field obtained numerically, we obtained the distribution of reflected light virtually, which was in agreement with the experimentally obtained image.
The proposed procedure may enable the one-to-one correspondence between the contrast image of spiral wave obtained experimentally and the velocity field of the spiral state obtained numerically.
For future research, the present procedure to compare experimental and numerical images may be extended so as to distinguish circular or spiral-shaped states identified in previous studies on spherical Couette flow.

\begin{flushleft}
{\large \bf Acknowledgement}  
\end{flushleft}
The authors would like to thank Mr. Kazuki Ota, Ms. Saki Tsumura, and Mr. Fumitoshi Goto for the pilot survey of the draft.
We would also like to thank Editage (www.editage.com) for English language editing.
This work was supported in part by the Grant-in-Aid for Scientific Research(C), JSPS KAKENHI, Grant No. 20K04294.
This study also benefited from the interaction within RISE-2018 No. 824022 ATM2BT of the European Union Horizon 2020-MSCA program, which includes Kansai University.


\bibliographystyle{spphys}       
\bibliography{scf11}

\end{document}